# *Plasmonics without negative dielectrics*

*Cristian Della Giovampaola and Nader Engheta**

*University of Pennsylvania*
*Department of Electrical and Systems Engineering*
*Philadelphia, Pennsylvania 19104, USA.*
*\*e-mail: engheta@ee.upenn.edu (corresponding author)*

**Abstract**

Plasmonic phenomena are exhibited in light-matter interaction involving materials whose real parts of permittivity functions attain negative values at operating wavelengths. However, such materials usually suffer from dissipative losses, thus limiting the performance of plasmon-based optical devices. Here, we utilize an alternative methodology that mimics a variety of plasmonic phenomena by exploiting the well-known structural dispersion of electromagnetic modes in bounded guided-wave structures filled with only materials with positive permittivity. A key issue in design of such structures is prevention of mode coupling, which can be achieved by implementing thin metallic wires at proper interfaces. This method, which is more suitable for lower frequencies, allows designers to employ conventional dielectrics and highly conductive metals for which the loss is low at these frequencies, while achieving plasmonic features. We demonstrate, numerically and analytically, that this platform can provide surface plasmon polaritons, local plasmonic resonance, plasmonic cloaking and epsilon-near-zero (ENZ)-based tunneling using conventional positive-dielectric materials.

## I. INTRODUCTION

Electromagnetic and optical signals can be manipulated and controlled by materials. The light-matter interaction in general exhibits temporal (i.e., frequency) and spatial (i.e., wave vector) dispersions. The frequency dispersion can be due to material dispersive response, which is the result of temporal characteristics of the polarization and magnetization densities in the materials. Such dispersions, e.g., the well-known Drude and Lorentzian dispersions, may endow some materials with negative real part of permittivity in certain frequency regimes, such as noble metals in the infrared and visible wavelengths. These epsilon-negative (ENG) materials play an important role in the fields of plasmonic optics and photonic metamaterials that witness growing interest in recent years.[1–5] The "natural" temporal dispersion of a material is not the only type of dispersion in wave interaction with materials. There is another form of dispersion, known as the geometrical or structural dispersion, which is due to the role of geometry of the structures on wave propagation, such as the modal dispersion in bounded guided-wave structures.[6,7] The parallel-plate waveguides (PPW) and the rectangular waveguides (RW) made of perfectly electric conducting (PEC) walls have been known for their structural dispersions, allowing the possibility to define an "effective" permittivity for a desired mode in such structures. These effective permittivity functions depend on the geometric dimensions of the guided-wave structure, filling materials, and the operating frequency. As an example, it is known that if one considers a $TE_{10}$ mode propagating in a PPW with PEC ceiling and floor, one can express the relative effective permittivity of the TE10 mode as[6,7] $\varepsilon_{eff} = \varepsilon_b - \lambda_0^2/4a^2$, where $\varepsilon_b$ is the relative permittivity of the medium filling the waveguide, $a$ is the separation between the parallel metallic plates, and $\lambda_o$ is the free-space wavelength (Fig. 1(a)). According to the above equation, one can design the PPW such that the effective permittivity of the mode is positive (effective-double-positive media, EDPS), zero (effective-epsilon-near-zero, EENZ), or negative (effective-epsilon-negative-media, EENG). If one considers the field distribution for the $TE_{10}$ mode (Fig. 1(a)), one will notice that in the middle mathematical plane parallel with the metallic plates of the PPW, both the electric and the magnetic fields have only transverse components with respect to the direction of propagation of this $TE_{10}$ mode (the longitudinal component of the magnetic field $H_y$, although non-zero inside this waveguide, vanishes on this middle plane). Therefore, if we restrict our observation to the middle plane of the PPW, the electromagnetic field distribution resembles the field distribution of a plane wave, which appears to propagate in an "effective" medium whose permittivity is equal to the effective permittivity of the mode inside the PPW, as expressed in the aforementioned equation. It is worth highlighting the fact that this effective permittivity exhibits the Drude dispersion, even if the filling material may be nondispersive, e.g., free space. Since this dispersion is strongly related to waveguide structures, the resulting effective media can be called *waveguide metamaterials*.

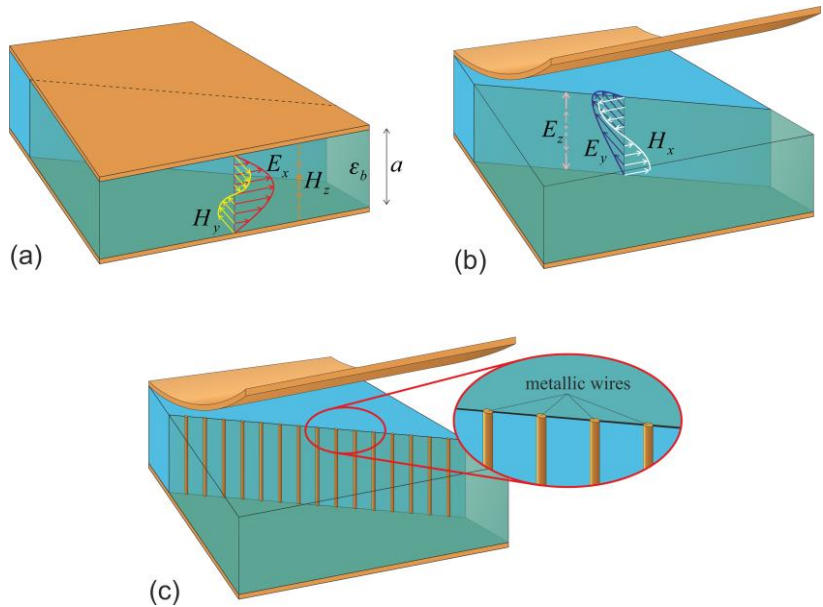

FIG. 1. A parallel plate waveguide (PPW) with the $TE_{10}$ mode, exhibiting specific effective permittivity tailored by the geometric dispersion. (a) The PPW height *a* and the relative permittivity of the filling medium $\varepsilon_b$ determine the effective permittivity of the propagating $TE_{10}$ mode. The field distribution of the $TE_{10}$ mode inside the PPW shows that the electric field is always orthogonal to the direction of propagation (the *y* axis in this case). We also note that the longitudinal component of the magnetic field vanishes on the middle mathematical plane of the PPW. (b) Field distribution of the $TM_{10}$ mode that may be generated at the interface between two media inside the PPW, due to the mode coupling from $TE_{10}$ to $TM_{10}$. (c) Thin metallic wires parallel to the *z* axis may short-circuit the electric field $E_z$ of the $TM_{10}$ mode, thus preventing coupling from the dominant $TE_{10}$ mode to the undesirable $TM_{10}$ mode.

The advantage of this approach, which is entirely different from the Mie-resonance-based all-dielectric scenarios studied in recent years (e.g., Ref. 8–10), is that an effectively positive, zero or negative permittivity for the $TE_{10}$ mode may, in principle, be achieved using PPW waveguide structures filled with conventional dielectric materials, e.g., air. In particular, in the microwave and THz regime this can be an efficient way to mimic epsilon-negative (ENG) and epsilon-near-zero (ENZ) media, which are not easily available in these frequency bands. Moreover, as an added bonus, one can avoid or reduce the deteriorating effects of losses in such effective ENG structures by choosing low-loss conventional dielectric materials with positive relative permittivity as filling materials and having metals such as copper for the walls in such guided-wave structures noting that at these frequencies metals are practically impenetrable and thus do not exhibit noticeable loss. There are studies[11,12] on waveguides below the cutoff as a way of providing effective negative permittivity or permeability but their focus is primarily on the propagation of either TE or TM modes inside such waveguides with the help of split-ring resonators or other inclusions in order to achieve propagation characteristics of left-handed media. Unlike these mentioned works, here we extend the study to a broader variety of effects, which involve pairing of media with positive and negative effective permittivity. In the present work, we analytically and numerically demonstrate that the analogy between media with actual positive, zero or negative permittivity and their counterparts where the permittivity is only *effectively* negative, zero or positive holds true not only for the propagation constants of conventional propagating scenarios, but also for several aspects of wave-structure interactions that may support some exotic phenomena, such as surface plasmon polaritons (SPP) surface-wave propagation and local surface plasmon resonance that are usually observed in the infrared and optical regimes, scattering-cancellation cloaking and ENZ-based supercoupling, just to name a few.

Although such waveguide structures may be an interesting way to mimic the ENG materials, such simple geometrical dispersion provided by the PPW is not sufficient for achieving some of the interesting phenomena in wave interaction with combinations of ENG and double-positive (DPS) materials. In fact, at the interfaces between different filling materials inside the PPW, mode coupling between the $TE_{10}$ and the $TM_{10}$ modes may occur, complicating the notion of the effective material parameters. It is important to highlight that the modal dispersion we mentioned above in order to mimic the effective material parameters is for a specific mode, namely, the $TE_{10}$ mode, and that the effective permittivity is determined when we have *only* this $TE_{10}$ mode. However, when the mode coupling is possible at the interface of two regions inside the PPW, the $TM_{10}$ mode can also be excited, since the $TM_{10}$ has the same cut-off frequency as the $TE_{10}$ mode in the PPW (the field distributions for $TM_{10}$ mode are shown in Fig. 1(b)), Therefore, due to such unwanted mode coupling, the fundamental $TE_{10}$ mode may not be the only mode available in the PPW, causing the structure not to exhibit the desired effective parameters that we desire. (As an aside, the higher order modes at the interfaces may also be generated due to such mode coupling, thus locally modifying the field distributions at such interfaces, even if those higher-order modes are below the cut-off conditions for that specific waveguide.) In order to prevent such mode coupling, remedy this shortcoming and therefore guaranteeing the fundamental $TE_{10}$ mode to be the only mode in the structure, we propose to place a set of very thin vertically-oriented metallic wires at the interface between different filling media in the PPW, directed perpendicular to the ceiling and floor of the PPW (Fig. 1(c)), allowing the $TE_{10}$ mode to be propagating (since its electric field is orthogonal to such thin metallic wires), while "short-circuiting" and suppressing the $TM_{10}$ mode which has a component of the electric field (i.e., $E_z$) parallel to these wires (Fig. 1(b)).

## II. WAVEGUIDE METAMATERIALS

In order to examine how the wire-added waveguide metamaterials may exhibit propagation characteristics dictated by the effective permittivity achieved through geometric dispersion inside the waveguides, we numerically study some of the plasmonic phenomena using only conventional DPS dielectrics inside PPW. In other words, we explore how we can achieve plasmonic features without using any natural ENG media. In all the following examples, without loss of generality, the operating frequency, just as an example, is set to $f_0 = 300 \text{ MHz}$, corresponding to the free-space wavelength of $\lambda_0 = 1$ m. At this frequency, it is expected that all materials, i.e., positive-permittivity and metallic materials we use in our numerical simulations to be effectively lossless.

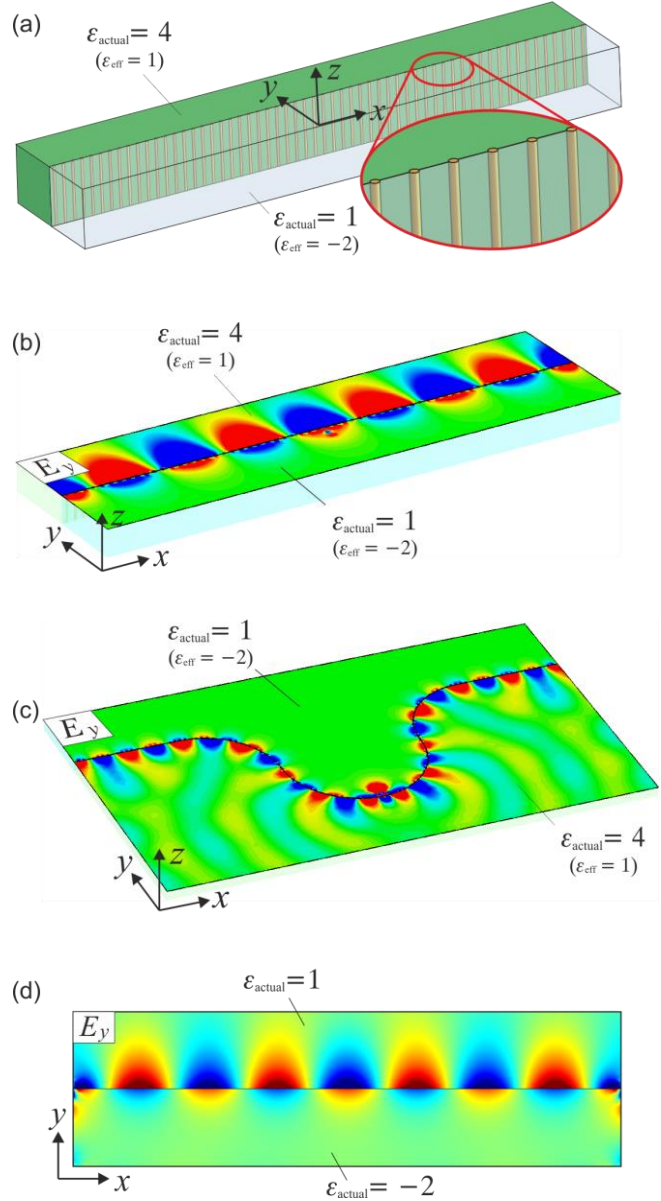

FIG. 2. Surface plasmon polariton (SPP) at the interface between two conventional (i.e., positive-permittivity) dielectric materials. (a) View of the interface inside the PPW. The top and bottom metallic plates of the waveguide are not shown here for easy observation. A series of thin metallic wires are located at the interface in order to short-circuit the $TM_{10}$ mode. (b) Snap shot value of the *y*-component of the electric field revealing an SPP-like mode propagating along the interface between the two media and exponentially decaying in the direction orthogonal to it. Furthermore, the sign of the *y*-component of the E field flips from one side of the interface to the other, analogous to the SPP propagation characteristics. (c) SPP along an arbitrarily curved interface between positive-permittivity media with the same effective permittivity as for case (b). The thin metallic wires are located all along the curved interfaced between the two media. (d) Conventional SPP generated in the ideal case of two natural media with true relative permittivity of 1 and -2.

As the first example, we analyze an interface between an EDPS and an EENG layer inside a PPW (Fig. 2(a)) in order to establish whether such structure can support a surface plasmon polariton (SPP) surface wave.[13] It is well known that when a natural material with actual positive permittivity is paired with another natural material whose actual permittivity has a negative real part, SPPs can propagate along the interface while being strongly confined in the transverse direction (i.e. normal to the interface). But here the structure we simulate consists of two slabs of positive dielectric with relative permittivity $\varepsilon_1 = 4$ and $\varepsilon_2 = 1$ (i.e., air) placed inside a parallel plate waveguide whose metallic plates are separated by a distance of about $a = 0.29\lambda_0$ so that the effective relative permittivity of the two layers turns to be $\varepsilon_{eff1} = 1$ and $\varepsilon_{eff2} = -2$. In Fig. 2(a) we place a series of thin metallic wires (of radius $\lambda_0/200$) along the entire interface, separated by $d = \lambda_0/20$ from each other. A short electric dipole, located at a distance of $\lambda_0/20$ from the interface and orthogonal to it, is used to excite the structure. In Fig. 2(b) a snap shot of the *y*-component of the electric field $E_y$, on the middle plane of the PPW, and normal to the interface is reported (for the sake of brevity, we only show the *y*-component of the electric field here). It is known that the wavelength of a conventional SPP propagating between two natural media with permittivity $\varepsilon_{m1}$ and $\varepsilon_{m2}$ is $\lambda_{SPP} = \lambda_0 \sqrt{(\varepsilon_{m1} + \varepsilon_{m2})/\varepsilon_{m1}\varepsilon_{m2}}$.[2,13] The simulated field distribution in our waveguide structures in Fig. 2(b) reveals a surface wave running along the interface with wavelength of $0.709\lambda_0$, which is consistent with the theoretical wavelength $\lambda_{SPP} = 0.707\lambda_0$ that an SPP would have if propagating along the interface of two semi-infinite media with actual relative permittivity values equal to the effective relative permittivities $\varepsilon_{eff1}$ and $\varepsilon_{eff2}$ of the two slabs in Fig. 2. Furthermore, the SPP generated at the EDPS-EENG interface has an exponential decay along the *y* axis and the component of the field normal to the interface flips sign across the interface, consistently with the propagation of a "natural" SPP. It is worth highlighting the fact that the nature of the SPP propagating along the interface in the waveguide metamaterials is completely different from the mechanism of the spoof plasmons,[14,15] since the metal wires do not contribute to the generation of the SPP but only to elimination of the $TM_{10}$ mode (a proof of this statement is provided through additional results in the Appendix A. We note that these thin wires also provide a platform for electric charges to accumulate in order to sustain the normal components of the electric field to face opposite to each other at the interface with wires). An additional example that shows the propagation characteristics of the SPP generated between the EDPS-EENG interface is shown in Fig. 2(c), where the boundary between the two positive-permittivity materials (with the same effective permittivity of the above example) is curved and composed of pieces of circles with radius of curvature of $\lambda_0$. We note that the SPP follows the curved path with no attenuation along it, whereas its variation in the transverse direction, i.e., in the direction normal to the local tangent of the interface is exponential. Finally, Fig. 2(d) shows the ideal case of an SPP propagating along a straight interface between two natural media with actual relative permittivity 1 and -2. There is a strong resemblance to the example of Fig. 2(b), thus confirming that the nature of the surface wave is consistent with the effective permittivity of the materials inside the PPW, while the two filling materials have positive permittivity, and does not depend on the (arbitrary) location of the wires.

Another well-known plasmonic phenomenon is the plasmonic resonance of a particle with negative permittivity whose size is much smaller than the operating wavelength.[16] Specifically, we investigate the possibility of mimicking the resonance of a cylindrical structure with negative effective permittivity $\varepsilon_{eff2}$ when embedded in a medium with effective permittivity $\varepsilon_{eff1} > 0$, again using regular DPS dielectrics for both the cylinder and the bulk medium. The cylindrical rod has radius $\lambda_0/30$ and it is placed inside a rectangular waveguide (Fig. 3(a)).

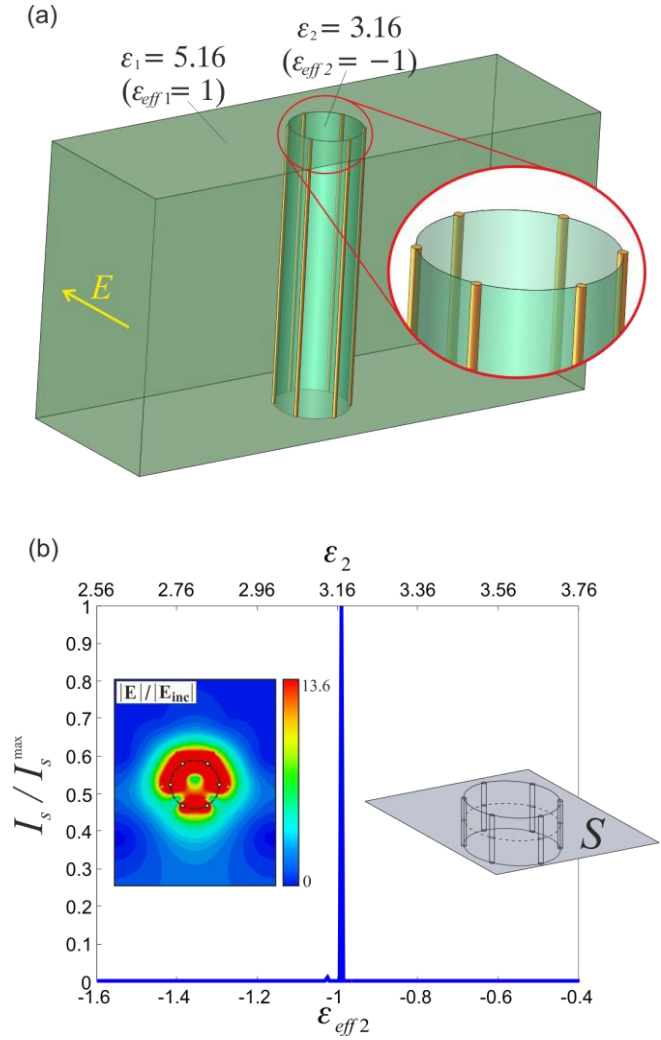

FIG. 3. Resonant scattering with all-positive-dielectric materials inside a rectangular waveguides. (a) Cylinder of relative permittivity $\varepsilon_2 = 3.16$ embedded into a bulk medium of permittivity $\varepsilon_1 = 5.16$ and surrounded by thin metallic wires. According to the dispersion relation of the waveguide, the effective relative permittivity of the cylinder and the surrounding medium become $\varepsilon_{eff2} = -1$ and $\varepsilon_{eff1} = 1$, respectively. (b) Normalized integral of the electric field amplitude (blue line) over a square surface $S$ crossing the cylinder at the middle plane of the waveguide and orthogonal to its axis, for different values of the cylinder relative permittivity. In the insets, the total electric field magnitude (normalized with respect to the incident field) and the position of the integration surface (on the right) are shown. We note that at the value of cylinder relative permittivity of 3.16 (when its effective relative permittivity becomes -1), we get resonant scattering.

The waveguide is filled with dielectric with relative permittivity $\varepsilon_1 = 5.16$ whereas the cylinder is made of dielectric with relative permittivity $\varepsilon_2 = 3.16$. Owing to the geometric dispersion of the waveguide (whose height was set to be $a = 0.245\lambda_0$), the effective values for the relative permittivity of the bulk medium and of the cylinder become $\varepsilon_{eff1} = 1$ and.., respectively, with the condition $\varepsilon_{eff2} = -\varepsilon_{eff1}$ being the requirement for the plasmonic resonance of an electrically small long cylindrical structure. Similarly to the structure described in Fig. 2(a), also here a set of 6 thin metallic wires of radius $\lambda_0/400$ have been inserted on the cylinder's outer surface which in this case is the interface between the EENG and the EDPS media, in order to short-circuit the undesired TM modes and thus prevent the mode coupling. In order to quantify the scattering enhancement and to prove that the scattering is maximum when the condition $\varepsilon_{eff2} = -\varepsilon_{eff1}$ is met as predicted by the theory, we numerically conduct a parametric study for different values of relative permittivity of the cylinder (which also corresponds to vary its effective relative permittivity in the same fashion, since both the height of the waveguide and the frequency are kept constant) and for each value of $\varepsilon_2$ (or its equivalently effective value $\varepsilon_{eff2}$) we compute the integral $I_s$ of the electric field magnitude over a square cross sectional surface $S$ perpendicular to the cylinder axis and lying on the waveguide middle plane. Fig. 3(b) shows the location of the surface $S$ with respect to the cylinder, and also the normalized value of $I_s$ as a function of $\varepsilon_2$. The peak of $I_s$ in the plot of Fig. 3(b) occurs at $\varepsilon_2 = 3.17$ (or equivalently at the effective value of $\varepsilon_{eff2} = -0.99$), thus confirming that the "effective" plasmonic resonance of the cylindrical structure is caused by the proper value of effective permittivity of the cylinder $\varepsilon_{eff2}$, and not by the wires or other factors (see the Appendix B for additional results providing evidence for our conclusion about the effect of the wires). The amplitude of the electric field vector at the middle plane of the waveguide when $\varepsilon_{eff2} = -0.99$ is shown in the inset in Fig. 3(b), where we notice a clear enhancement of the electric field surrounding the cylinder.

A similar structure we now use to verify another interesting application of plasmonic materials that has been proposed in the past, i.e., the plasmonic cloak.[17] In this technique, the scattering cross section of either a dielectric object can be reduced by coating the object with a plasmonic cover made of material with relative permittivity smaller than 1, or with negative values. By properly designing the ratio between the object and the cloak sizes, the scattered field produced by the cloak tends to cancel out the field scattered by the object, thus decreasing the overall scattering signature of the initial object. We now show that with our structure we can achieve similar effect using only positive-permittivity materials. We consider a cylinder with radius $r_c = \lambda_0/30$ and actual relative permittivity $\varepsilon_2 = 6.16$ embedded in a rectangular waveguide filled with a medium with relative permittivity $\varepsilon_1 = 5.16$, and a cylindrical cloak surrounding the cylinder with relative permittivity $\varepsilon_3 = 4.54$ (Fig. 4(a)). Since the height of the waveguide was set to be $a = 0.245\lambda_0$, the effective values of relative permittivity of the different parts of the structures turned out to be $\varepsilon_{eff1} = 1$ for the bulk medium, $\varepsilon_{eff2} = 2$ for the cylinder and $\varepsilon_{eff3} = 0.38$ for the cloak. According to Ref. 17, with the above configuration, the cloak radius $r_{cl}$ is required to be about $r_{cl} = 1.25 r_c$. Also, a set of 10 thin metallic wires and additional 15 wires (all of radius $\lambda_0/1000$) are placed between the cylinder and the cloak, and at the outer surface of the cloak, respectively, as shown in Fig. 4(a). The amplitude of the scattered fields for the cloaked and the uncloaked cylinder are evaluated and compared, and are reported in Fig. 4(b), where we notice an overall reduction of the scattered field when the "effective" plasmonic cloak is used.

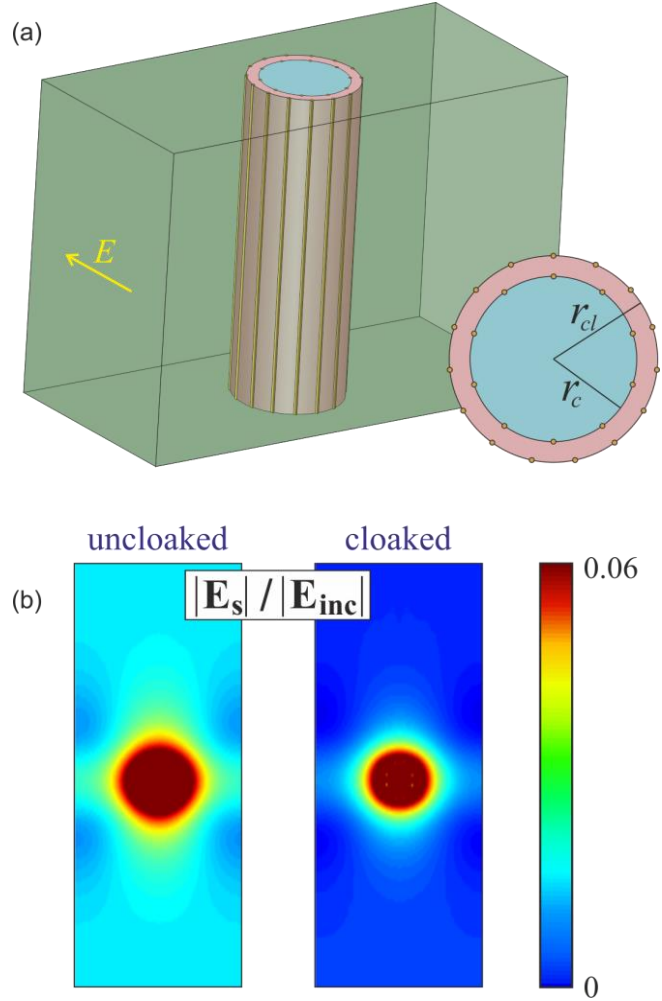

FIG. 4. Plasmonic cloaking with all-positive-dielectric materials inside a rectangular waveguides. (c) Dielectric cylinder surrounded by a dielectric layer, which acts as a plasmonic cloak when considering the effective values of the relative permittivity inside the waveguide. Both the cylinder and the cloak are surrounded by thin metallic wires that suppress the $TM_{10}$ mode. (d) Amplitude of the electric scattered field (normalized with respect to the incident field) for the uncloaked (left) and the cloaked (right) cylinder, showing a significant reduction of the scattering signature of the cylinder when the cloak is applied. We emphasize that all materials in our simulations are with positive permittivity.

As a final example, we study the application of our method to the phenomenon of supercoupling due to the epsilon-near-zero (ENZ) structures. It is well known that whenever geometric changes, such as bending or squeezing, occur in a conventional waveguide, the propagating mode may be partially reflected. In Ref. 18–22 it was shown that when the junction between two waveguides is filled with an ENZ material,[18–22] the incoming energy may, under certain conditions, tunnel through the junction with effectively no reflection even though the junction may be significantly bent and squeezed. We now show that this effect can be obtained with our method using only positive-permittivity materials. We design two rectangular metallic waveguides connected by a narrow channel. The waveguides are filled with a dielectric with relative permittivity $\varepsilon_b = 5$, while the channel is filled with another dielectric with relative permittivity $\varepsilon_c = 4$. The height of the waveguide is set to be $a = \lambda_0/4$, causing the effective relative permittivity of these two regions to dispersion, the effective relative permittivity for the bulk and the channel are 1 and 0, respectively, and thus the ENZ supercoupling occurs at our design frequency. The channel length is $l_c = 0.6\lambda_0$. In the reflection/transmission plot (*s*-parameters, $s_{11}$ and $s_{21}$) underneath we notice two transmission peaks; the first peak is associated with the ENZ tunneling effect and it occurs at the design frequency, whereas the second peak is due to a Fabry-Perot effect. (b) Waveguide with ENZ channel similar to the configuration described in (a), but with a longer channel $l_c = 0.9\lambda_0$, revealing the fact that while the transmission peak due to the Fabry-Perot effect exhibits a red shift as expected, the ENZ tunneling remains at the design frequency. (c) The waveguide has the same channel length as in (a), but the channel is now sharply bent at $90^0$. The reflection/transmission characteristics do not change, showing again one of the characteristics of ENZ supercoupling. (d) All-air-filled waveguides with narrow channel. The channel is achieved with a waveguide whose height is reduced with respect to the feeding arms of the waveguides, in order to achieve an effective relative permittivity close to zero in the channel. The *s*-parameters of the waveguide show a transmission peak at roughly the design frequency. (e) Configuration similar to (d), but with a longer channel in order to show that the ENZ tunneling stays at the design frequency.

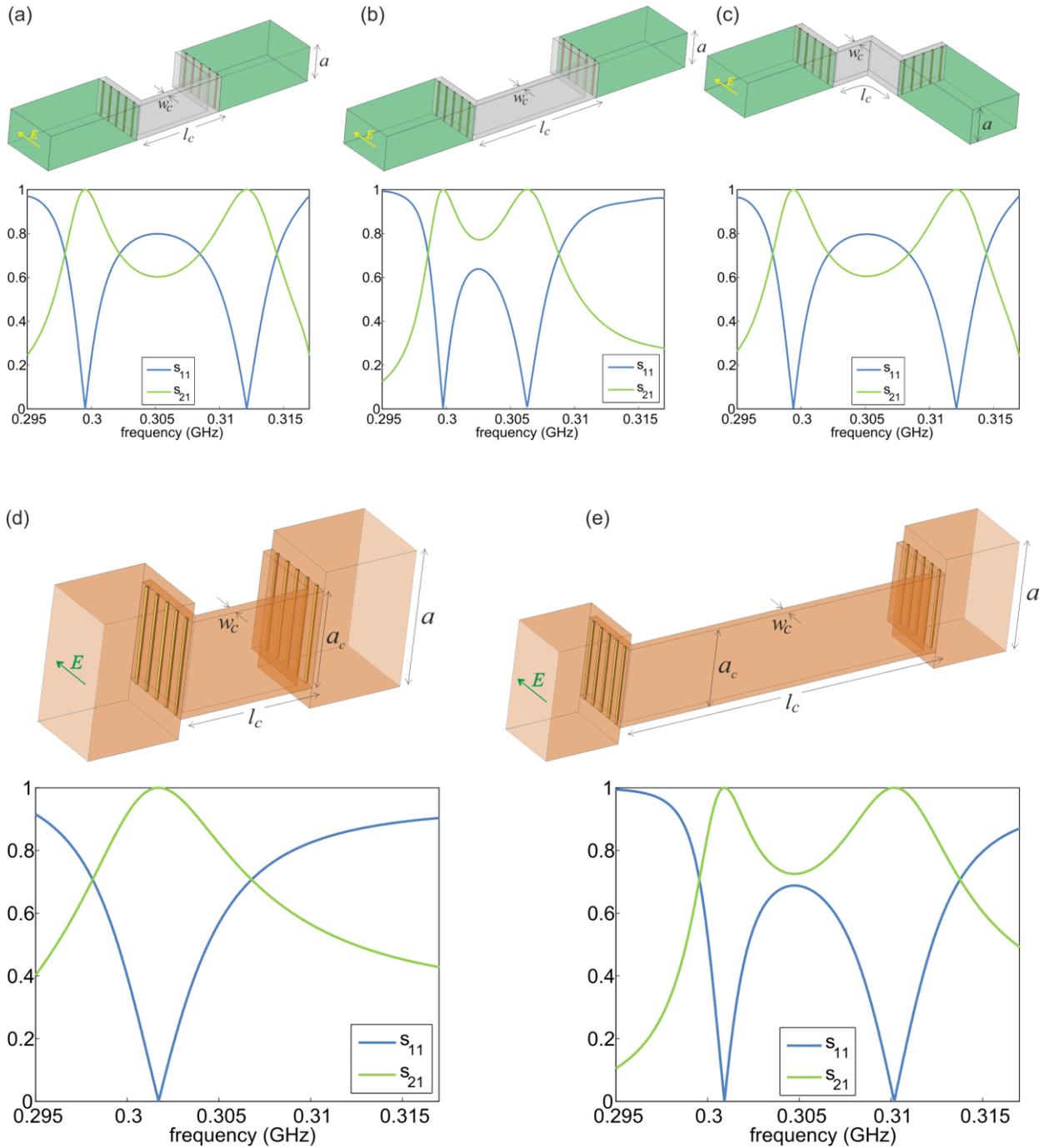

FIG. 5. Epsilon-Near-Zero (ENZ) supercoupling with all positive-dielectric materials. (a) Two waveguides filled with a dielectric medium with relative permittivity $\varepsilon_b = 5$, connected via a channel filled with a dielectric with relative permittivity $\varepsilon_c = 4$. Owing to the waveguide structural

become $\varepsilon_{eff,b} = 1$ and $\varepsilon_{eff,c} = 0$, respectively, due to the structural dispersion we discussed. In all the numerical results shown in Fig. 5, the waveguides are fed by a $TE_{10}$ mode with the electric field orthogonal to the thin metallic wires. Furthermore, the width of the channel is $w_c = \lambda_0 / 20$ and the waveguide width (along the electric field direction) is $0.4\lambda_0$ for all the structures under study. Figure 5(a) shows the two waveguides separated by the narrow channel. The metallic wires are located at the interface between the waveguide bulk and the beginning of the channel, which has a U-shape.[18–21] In this configuration, the length of the channel is ... The plot underneath the geometry, in Fig. 5(a), shows the reflection and transmission characteristics of the waveguide for the $TE_{10}$ mode as a function of frequency. We notice two transmission peaks; the first peak is associated with the effective ENZ tunneling effect and it occurs at our design frequency of 300 MHz, whereas the second transmission peak at a higher frequency is due to a Fabry-Perot effect, since at that frequency the effective permittivity of the channel is positive and the effective wavelength inside the channel is now finite. In order to confirm the difference between the two mechanisms behind the two transmission peaks, we simulate the same waveguide with a longer channel of $l_c = 0.9\lambda_0$ (Fig. 5(b)). By looking at the reflection/transmission plot, we observe a red shift in the Fabry-Perot as expected since the channel length has been increased. However, the ENZ tunneling peak remains at the design frequency, since at this frequency the effective wavelength inside the channel is in principle infinitely long. Furthermore, we also investigate the effect of a possible tight bending of the channel, and we simulate the structure depicted in Fig. 5(c), where the channel has the same length as in Fig. 5(a), but it now has a 90º sharp bend. As predicted by the ENZ tunneling phenomenon, the reflection/transmission properties of this structure at the ENZ frequency are identical to the ones shown in Fig. 5(a), since the ENZ condition is not affected by the shape of the channel (provided that its longitudinal cross-sectional area is kept electrically small). As the next step, we remove any dielectric medium from inside the waveguides and the channel, and consider all-air-filled structures where the only material used is PEC for all walls and the thin wires. Since the air is the only material inside the waveguide, the change in effective permittivity for the waveguide and the channel can only be achieved by considering different heights of these structures. In the examples shown in Figs. 4D and 4E, the height of the two waveguides is $a = 0.7\lambda_0$ and the height of the channel is $a_c = 0.5\lambda_0$, thus resulting in effective relative permittivity values of 0.49 and 0 for the waveguide and the channel, respectively. The plot in Fig. 5(d) shows that even for the all-air-filled case of our structure, the ENZ tunneling still happens around the design frequency, thus confirming that all these phenomena are dictated only by the effective permittivity resulting from the structural dispersion. Finally, Fig. 5(e) shows a version of the previously described all-air waveguide with a longer channel, in order to show the red shift of the Fabry-Perot related peak, different from the ENZ tunneling. As an aside, it is worth noting that it might be possible that in some rare instances the thin metallic wires may not be needed. For example, in the last example related to the ENZ tunneling shown in Fig. 5, the same structures would work even without the metallic wires, as in Ref. 20. However, this is a rare instance in which there is only one component of the electric field, which is orthogonal to the channel and consequently parallel to the waveguide-channel interface, therefore the mode coupling to $TM_{10}$ mode would not occur. We reiterate that in most cases the thin wires play the major role in preventing the mode coupling, thus essential in exploiting our structural dispersion in various plasmonic phenomena.

## III. CONCLUSIONS

In conclusion, we have demonstrated, theoretically and through a series of numerical simulations, that the structural dispersion of waveguides, when augmented with presence of thin metallic wires judiciously located at interfaces, may provide us with a useful platform, particularly in the microwave, millimeter-wave and THZ regimes, to demonstrate a variety of plasmonic phenomena, which in the infra-red and optical wavelengths can be done with natural materials with negative or near zero permittivity that may be lossy. However, since our proposed structures involve only positive-dielectric materials and metallic walls and wires, which exhibit very little loss at these lower frequencies, we can achieve various plasmonic features with low level of dissipation. This may open doors to new possibilities in design of low-loss metamaterials that offer exciting phenomena that have otherwise been limited by material loss.

**Acknowledgements**

This work has been supported in part by US Office of Naval Research (ONR) Multidisciplinary University Research Initiative (MURI) grant number N00014-10-1-0942, and by US Air Force Office of Scientific Research (AFOSR) Multidisciplinary University Research Initiative (MURI) grant number FA9550-14-1-0389.

**APPENDIX A**

The effect of the thin metallic wires mentioned in the main text has been investigated with additional simulations shown here. We simulated the same structure described in the main text and depicted in Fig. 2 for the surface plasmon polariton (SPP) generation without using the thin metallic wires at the interface between the two slabs of actual relative permittivity $\varepsilon_{actual1} = 4$ (so the effective relative permittivity $\varepsilon_{eff1}=1$) and actual relative permittivity $\varepsilon_{actual2}=1$ (so the effective relative permittivity of $\varepsilon_{eff2}=-2$). Although the two slabs have effective dielectric constants of opposite signs, Fig. A1 shows no excitation of any surface wave propagating at the interface between the two media. This is due to the fact that additional waveguide modes (the most significant one being a TM$_{10}$ mode) are generated at the interface, thus producing a field distribution that differs from the TE$_{10}$ mode for which the waveguide effect is considered in the initial assumptions. Moreover, the sign of the electric field normal to the interface does not change across the interface because the actual dielectric constants have the same signs. Fig. A2 reports the distribution of the electric field $E_z$ at the waveguide top plate and orthogonal to it, indicating that mode coupling from TE$_{10}$ mode to the TM$_{10}$ mode has indeed occurred. When the thin metallic wires are assumed at the interface, as shown in the main text the surface wave resembling the SPP becomes possible.

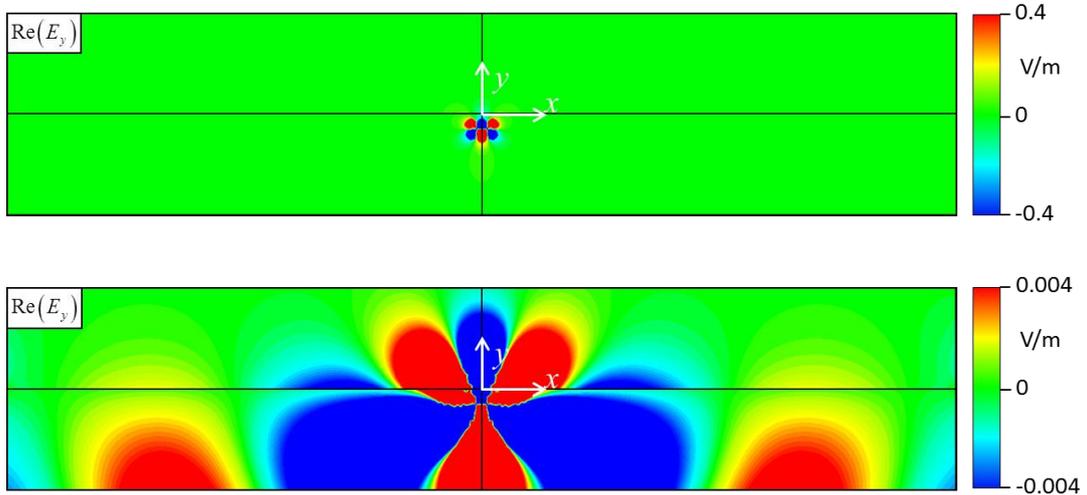

FIG. A1. Snapshot of the *y*-component of the electric field in V/m for a dielectric interface similar to the results of Fig. 2 in the main manuscript, but without using the thin metallic wires at the interface between the two slabs. The top plot has the same scale as the plot in Fig. 2. We clearly notice the absence of a surface wave along the interface. The bottom panel has an expanded scale in order to show the difference in the field distribution when compared with the field map of Fig. 2. We can also notice that the sign of the electric field $E_y$ does not change across the interface (since the actual permittivity of the two dielectric regions have the same signs), unlike the case where the metallic wires are used as shown in the main text.

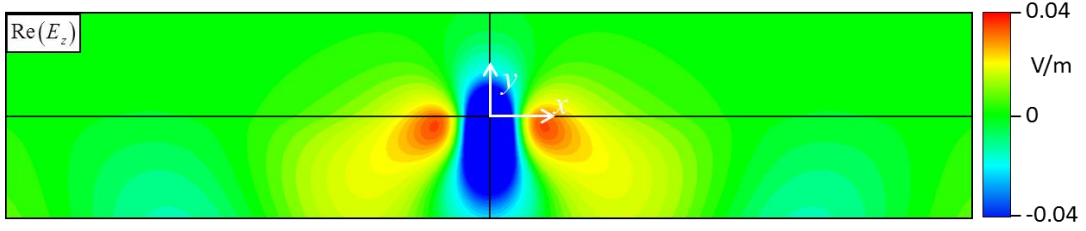

FIG. A2. Similar to Fig. A1, except here the plots shows the *z*-component of the electric field in V/m at the top plate of the waveguide. Owing to absence of the thin metallic wires, there is a significant non-zero *z*-component of the electric field at the top plate, indicating presence of the $TM_{10}$ mode due to mode coupling.

**APPENDIX B**

Similarly to what is described in the previous paragraphs, we also conducted a study on the effects of the thin metallic wires for the cylinder local "plasmonic" resonance. Fig. A3 shows the amplitude of the total electric field at the waveguide middle plane, normalized to the incident field. We do not notice any resonance phenomenon even if the effective permittivity of the cylinder is equal and opposite to the effective permittivity of the bulk medium. The absence of the resonance is due to the generation of the $TM_{10}$ mode (i.e., coupling of $TE_{10}$ mode to $TM_{10}$ mode) that is excited at the curved interface between the cylinder and the bulk. The generation of the $TM_{10}$ mode is verified by the presence of a significant *z*-component of the electric field at the waveguide top plate (the component that is orthogonal to the top and bottom plates), as depicted in the right panel of Fig. A3. In order to verify that the thin metallic wires by themselves do not play any role in the cylinder resonance, we numerically evaluated the field distribution for a cylindrical region surrounded by these thin wires located inside the waveguide, using the same medium for both the bulk and the cylinder (i.e., the actual materials inside and outside this cylindrical region are the same). From the field distribution in Fig. A4 we can conclude that the wires by themselves do not produce any "resonance" effect since the electric field at the middle mathematical plane in the waveguide has uniform amplitude distribution (left panel), meaning that the wires and the cylinder are invisible to the incident TE$_{10}$ mode. Moreover, the wires do not affect the TE$_{10}$ field distribution, since the $z$-component of the electric field at the top and bottom plates is zero as evident from the right panel of Fig. A4.

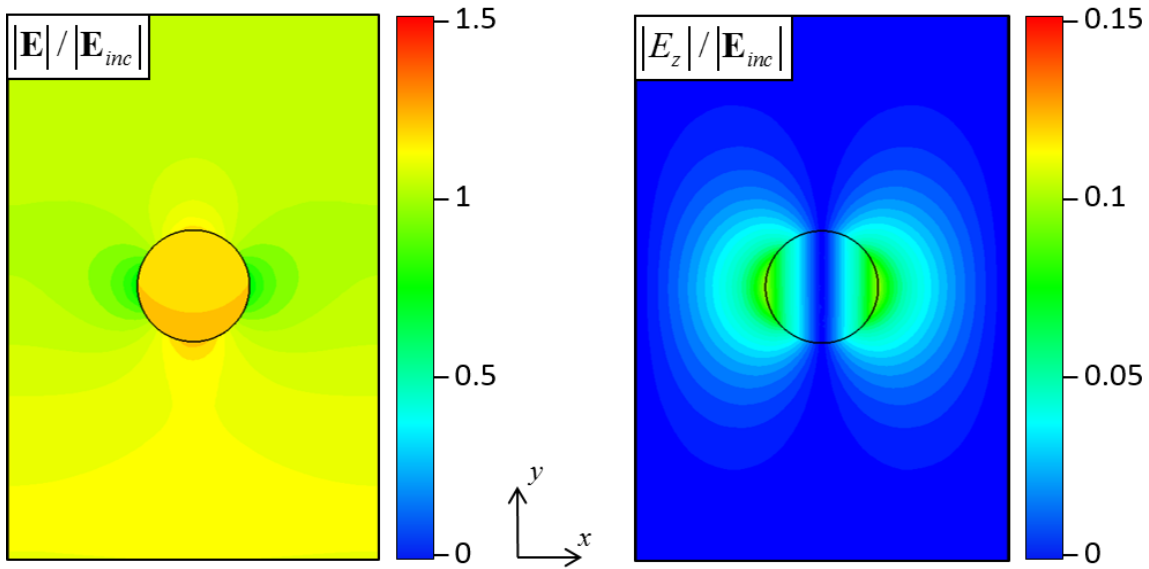

Fig. A3. Amplitude of the normalized total electric field at the middle plane of the waveguide (left) and of the normalized $z$-component of the electric field $E_z$ at the top plate of the waveguide (right), in the absence of the thin metallic wires surrounding the cylinder. The left plot shows the absence of any resonant phenomenon, whereas the right plot reveals that near the waveguide plates there is a significant value for the $z$-component of the electric field orthogonal to the metallic plates, implying the presence of the TM$_{10}$ mode generated due to coupling from TE$_{10}$ more to TM$_{10}$ mode at the interface between the bulk medium and the cylinder.

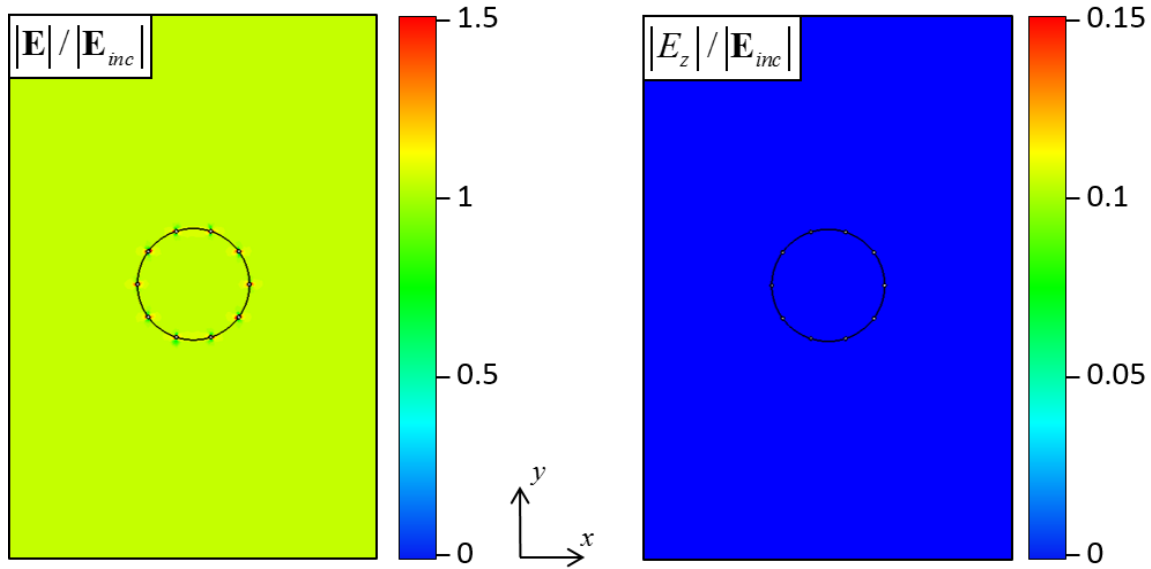

Fig. A4. Similar to Fig. A3, except here the material inside the cylinder and the material in the bulk are the same. The plot on the left shows the amplitude of the normalized total electric field on the middle plane of the waveguide being constant everywhere, indicating that there is no interaction between the bulk and the cylinder regions, as well as there is no effect produced by the metallic wires by themselves. The plot on the right reveals absence of any z-component of the electric field orthogonal to the top metallic plate since there is no material interface between the inside and outside the cylinder inside the waveguide and the metallic wires by themselves do not contribute to field perturbation.